\documentclass[preprint,12pt]{aastex}
\usepackage{graphicx}

\shortauthors{Winn et al.~2010}
\shorttitle{Spin-Orbit Alignment of HAT-P-13b}

\begin{document}

%
\def\ltsima{$\; \buildrel < \over \sim \;$}
\def\lsim{\lower.5ex\hbox{\ltsima}}
\def\gtsima{$\; \buildrel > \over \sim \;$}
\def\gsim{\lower.5ex\hbox{\gtsima}}
                                                                                          
%

\bibliographystyle{apj}

\title{ The HAT-P-13 Exoplanetary System:\\
Evidence for Spin-Orbit Alignment and a Third Companion }

\author{
Joshua N.\ Winn\altaffilmark{1,2},
John Asher Johnson\altaffilmark{3},
Andrew W.\ Howard\altaffilmark{4,5},
Geoffrey W.\ Marcy\altaffilmark{4},
G\'asp\'ar \'A.\ Bakos\altaffilmark{6,7},
Joel Hartman\altaffilmark{6},
Guillermo Torres\altaffilmark{6},
Simon Albrecht\altaffilmark{1},
Norio Narita\altaffilmark{2,8}
}

\altaffiltext{1}{Department of Physics, and Kavli Institute for
  Astrophysics and Space Research, Massachusetts Institute of
  Technology, Cambridge, MA 02139}

\altaffiltext{2}{Kavli Institute for Theoretical Physics, UCSB, Santa
  Barbara, CA 93106, USA}

\altaffiltext{3}{Department of Astrophysics, California Institute of
  Technology, MC~249-17, Pasadena, CA 91125}

\altaffiltext{4}{Department of Astronomy, University of California,
  Mail Code 3411, Berkeley, CA 94720}

\altaffiltext{5}{Townes Postdoctoral Fellow, Space Sciences
  Laboratory, University of California, Berkeley, CA 94720}

\altaffiltext{6}{Harvard-Smithsonian Center for Astrophysics, 60
  Garden St., Cambridge, MA 02138}

\altaffiltext{7}{NSF Fellow}

\altaffiltext{8}{National Astronomical Observatory of Japan, 2-21-1
  Osawa, Mitaka, Tokyo 181-8588, Japan}

\begin{abstract}

  We present new radial-velocity measurements of HAT-P-13, a star with
  two previously known companions: a transiting giant planet ``b''
  with an orbital period of 3 days, and a more massive object ``c'' on
  a 1.2~yr, highly eccentric orbit. For this system, dynamical
  considerations would lead to constraints on planet b's interior
  structure, if it could be shown that the orbits are coplanar and
  apsidally locked. By modeling the Rossiter-McLaughlin effect, we
  show that planet b's orbital angular momentum vector and the stellar
  spin vector are well-aligned on the sky ($\lambda = 1.9\pm
  8.6$~deg). The refined orbital solution favors a slightly eccentric
  orbit for planet b ($e=0.0133\pm 0.0041$), although it is not clear
  whether it is apsidally locked with c's orbit ($\Delta\omega =
  36_{-36}^{+27}$~deg). We find a long-term trend in the star's radial
  velocity and interpret it as evidence for an additional body ``d'',
  which may be another planet or a low-mass star. Predictions are
  given for the next few inferior conjunctions of c, when transits may
  happen.

\end{abstract}

\keywords{planetary systems --- planetary systems: formation ---
  stars:~individual (HAT-P-13) --- stars:~rotation}

\section{Introduction}

Precise radial-velocity measurements have revealed more than 30
multiple-planet systems (Wright~2010). However, in only a few cases
have transits been detected for any of the planets in those systems.
Those cases are potentially valuable because the transit
observables---the times of conjunction, orbital inclination, and
projected spin-orbit angle, among others---provide a much more
complete description of a planetary system, which may in turn give
clues about its formation and evolution. The Corot-7 system has two
orbiting super-Earths, one of which transits (L{\'e}ger et al.~2009,
Queloz et al.~2009). The HAT-P-7 system has a transiting hot Jupiter
in a polar or retrograde orbit, as well as a longer-period companion
that could be a planet or a star (P\'al et al.~2008, Winn et al.~2009,
Narita et al.~2010). The HAT-P-13 system, the subject of this paper,
features a G4 dwarf star with two previously known orbiting companions
(Bakos et al.~2009). The inner companion (HAT-P-13b, or simply ``b''
hereafter) is a transiting hot Jupiter in a 2.9 day orbit. The outer
companion (``c'') has an eccentric 1.2~yr orbit and a minimum mass
($M_c\sin i_c$) of about 15 Jupiter masses, although its true mass
($M_c$) and orbital inclination ($i_c$) are unknown. In particular,
transits of companion c have neither been observed nor ruled out.

Batygin, Bodenheimer, \& Laughlin (2009) and Mardling~(2010) showed
that it may be possible to use the observed state of the system to
determine planet b's Love number $k_2$, a parameter that depends on
the planet's interior density distribution. This would be of great
interest, as few other methods exist for investigating the interior
structure of exoplanets. The method is based on the theoretical
expectation that tidal evolution has aligned the apsides of the orbits
of b and c. This method has not yet yielded meaningful constraints on
$k_2$, partly because of the large uncertainty in the eccentricity of
b's orbit. Another relevant parameter is the mutual inclination
between the orbits, which is not known at all.

Radial-velocity observations are usually powerless to determine mutual
inclinations, unless the planets are in a mean-motion resonance (see,
e.g., Correia et al.~2010). However, for a transiting planet it is
possible to assess the alignment between the orbit and the stellar
equator through the Rossiter-McLaughlin (RM) effect. A system with
mutually inclined planetary orbits might also be expected to have
large angles between the orbits and the stellar equator. In
particular, Mardling (2010) presented a formation scenario for
HAT-P-13 involving gravitational scattering by a putative third
companion, which could have caused large mutual inclinations and a
large stellar obliquity.

In this paper we present new radial-velocity measurements of HAT-P-13
bearing on all these issues. The new data are presented in
\S~\ref{sec:rv}. Our analysis is presented \S~\ref{sec:analysis}, and
includes evidence for a third companion ``d'' (\S~\ref{subsec:third}),
refined estimates of the eccentricity and apsidal orientation of b's
orbit (\S~\ref{subsec:ecc}), modeling of the RM effect
(\S~\ref{subsec:lambda}), and updated predictions for the next
inferior conjunction (possible transit window) of companion c
(\S~\ref{subsec:dt}). In \S~\ref{sec:disc} we discuss the implications
for further dynamical investigations of HAT-P-13.

\section{Observations}
\label{sec:rv}

We observed HAT-P-13 with the High Resolution Spectrograph (HIRES;
Vogt et al.~1994) on the Keck~I 10m telescope, using the same
instrument settings and observing protocols that were used by Bakos et
al.~(2009) and that are used by the California Planet Search (Howard
et al.~2009). In particular, we used the iodine gas absorption cell to
calibrate the instrumental point-spread function and the wavelength
scale.  The total number of new spectra is 75, which are added to the
30 spectra presented by Bakos et al.~(2009). Of the new spectra, 40
were obtained on the night of 2009~Dec~27-28, spanning a transit of
HAT-P-13b, and were gathered to measure the RM effect. The other 35
were obtained on arbitrary nights. They extend the timespan of the
data set by approximately 1~yr, and thereby help to refine the orbital
parameters.

\begin{figure}[htb]
\epsscale{1.0}
\plotone{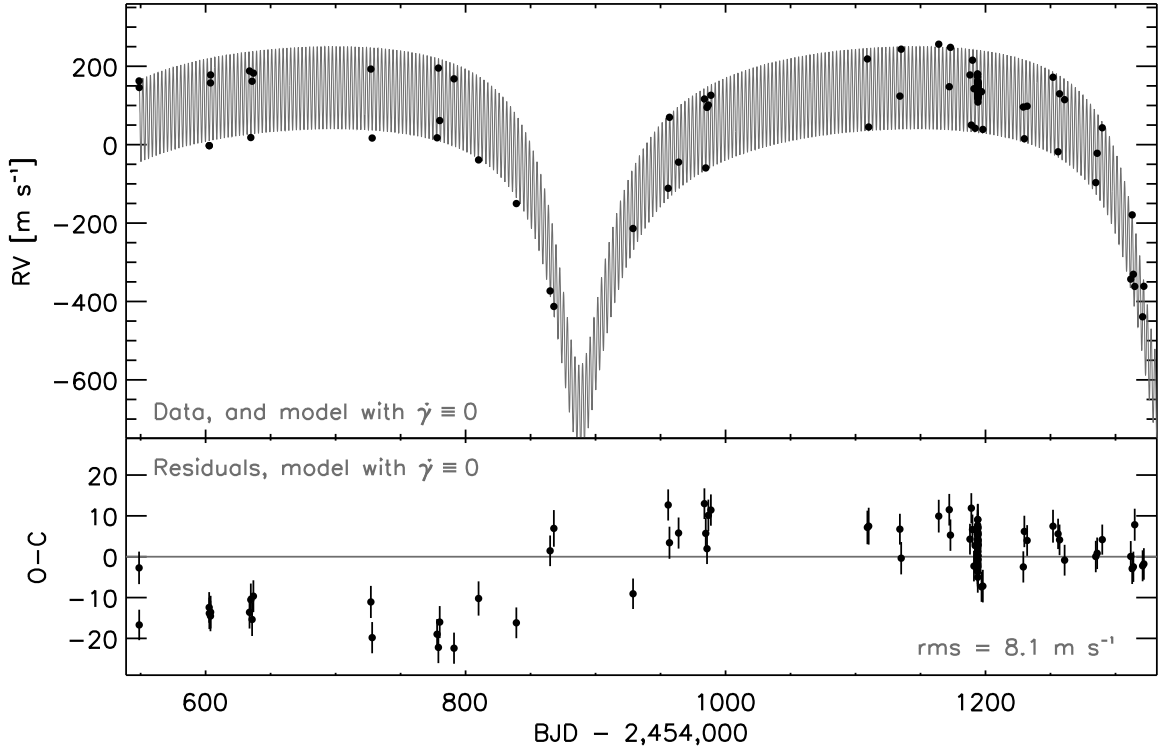}
\caption{{\bf Radial velocity variation of HAT-P-13.}  {\it
    Top.}---Measured RVs, and the best-fitting model. The model
  consisted of two Keplerian orbits and did not allow for any
  additional acceleration ($\dot{\gamma} \equiv 0$).  {\it
    Bottom.}---Residuals. The poorness of the fit, and the pattern of
  residuals, are evidence for a third companion.
\label{fig:rvtime1}}
\end{figure}

\begin{figure}[htb]
\epsscale{1.0}
\plotone{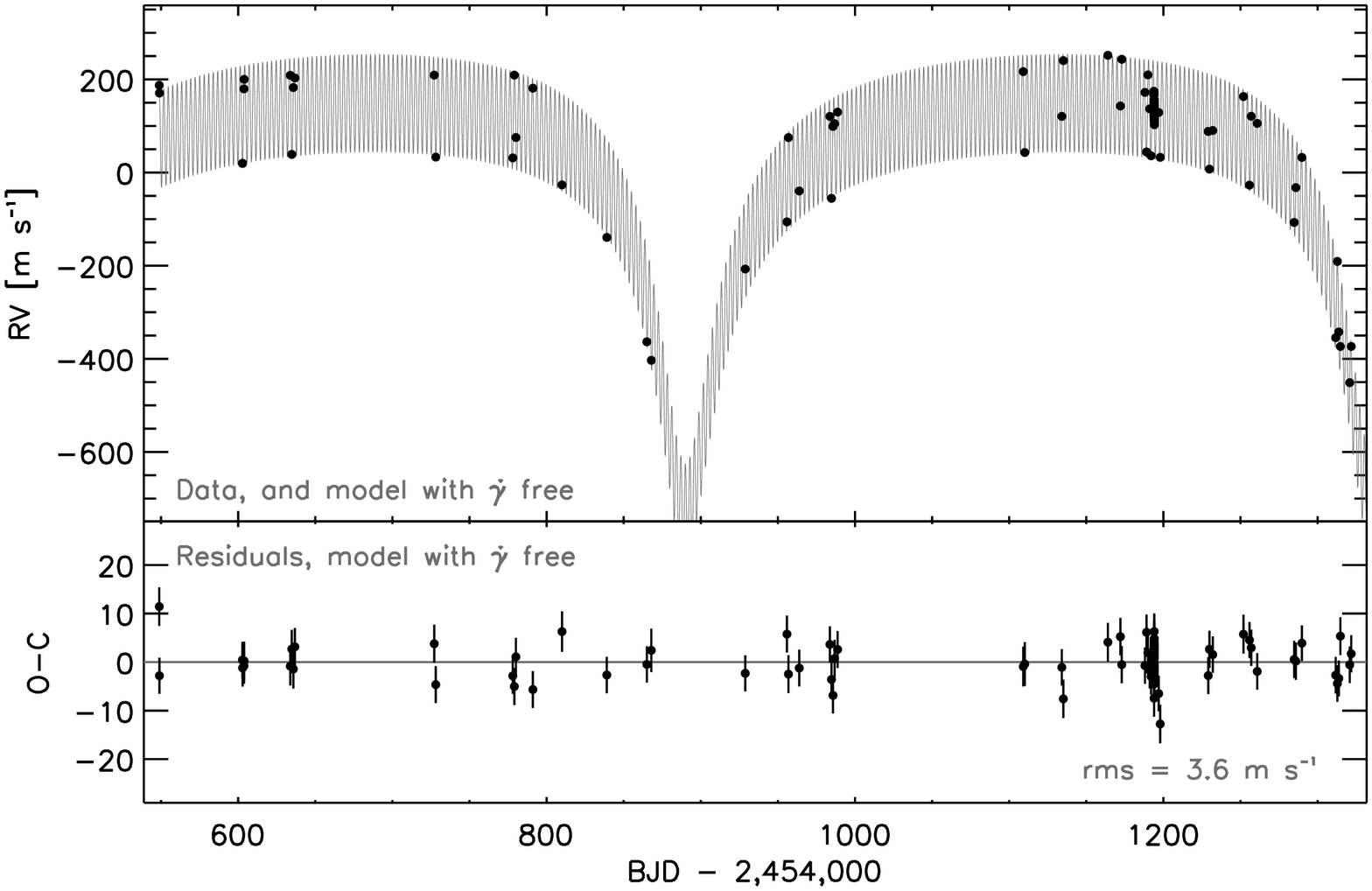}
\caption{{\bf Radial velocity variation of HAT-P-13.}
{\it Top.}---Measured RVs, and the best-fitting model, this time
  allowing for a constant radial acceleration ($\dot{\gamma}$) in
  addition to two Keplerian orbits. The best fitting value
  of $\dot{\gamma}$ was 17.5~m~s$^{-1}$~yr$^{-1}$.
{\it Bottom.}---Residuals.
\label{fig:rvtime2}}
\end{figure}

The radial velocity (RV) of each spectrum was measured with respect to
an iodine-free template spectrum, using the algorithm of Butler et
al.~(1996) with subsequent improvements. All of the spectra obtained
by Bakos et al.~(2009) were re-reduced, for consistency.  Measurement
errors were estimated from the scatter in the fits to individual
spectral segments spanning 2~\AA. The RVs are given in
Table~\ref{tbl:rv}, and plotted as a function of time in
Figures~\ref{fig:rvtime1}-\ref{fig:rvtime2}. The model curves
appearing in those figures are explained in \S~\ref{sec:analysis}.

\section{Analysis}
\label{sec:analysis}

Our model for the radial-velocity data took the form
\begin{equation}
  V_{\rm calc}(t) = V_b(t) + V_c(t) + V_{\rm RM}(t) + \gamma + \dot{\gamma}(t-t_0),
\end{equation}
where $V_{\rm calc}$ is the calculated RV, $V_b$ and $V_c$ are the
radial velocities of non-interacting Keplerian orbits, $V_{\rm RM}$ is
the transit-specific ``anomalous velocity'' due to the
Rossiter-McLaughlin effect (\S~3.3), and $\{\gamma,\dot{\gamma},t_0\}$
are constants. The first constant, $\gamma$, specifies the velocity
offset between the system barycenter and the arbitrary template
spectrum that was used to calculate RVs. The second constant,
$\dot{\gamma}$, allows for a constant radial acceleration, and was
included because models with $\dot{\gamma}=0$ did not fit the data
(\S~3.1). We interpret $\dot{\gamma}$ as the acceleration produced by
a newly-discovered long-period companion ``d''. The third constant,
$t_0$, is an arbitrary reference time that was taken to be the time of
the first RV datum (BJD~2,454,548.80650).

Our RM model was based on that of Winn et al.~(2005), in which
simulated spectra are used to calibrate the relation between the phase
of the transit and the measured radial velocity. For this case we used
the relation
\begin{equation}
\Delta V(t) = -(v\sin i_\star)~\delta(t)
\left[
0.9833 - 0.0356\left( \frac{v_p(t)}{v\sin i_\star} \right)^2
\right],
\end{equation}
where $v\sin i_\star$ is the sky-projected stellar rotation speed,
$\delta$ is the fractional loss of light, and $v_p$ is the radial
velocity of the portion of the stellar photosphere that is hidden by
the planet. To calculate $v_p$ we assumed that the stellar photosphere
rotates uniformly with an angle $\lambda$ between the sky projections
of the spin vector and the orbital angular momentum vector (see, e.g.,
Ohta et al.~2005 or Gaudi \& Winn 2007).

Since $\delta$ depends on the planet-to-star radius ratio
$R_p/R_\star$, orbital inclination $i$, and impact parameter $b_{\rm
  tra}$, all of which are more tightly bounded by observations of
photometric transits than by the RM effect, we simultaneously fitted a
composite $i'$-band transit light curve based on the photometric data
of Bakos et al.~(2009). For the photometric model we assumed a
quadratic limb-darkening law and used the analytic formula of Mandel
\& Agol (2002), as implemented by P\'al (2008). Because the
photometric data are not precise enough to constrain both of the
limb-darkening coefficients $u_1$ and $u_2$, we fixed $u_2\equiv
0.3251$, the value obtained by interpolating the Claret~(2004) tables,
and allowed $u_1$ to vary freely.\footnote{The result, $u_1=0.269\pm
  0.076$, was consistent with the tabulated value of 0.3068.} For the
RM model, we used a linear law with a fixed coefficient of 0.72, as
appropriate for the $V$-band, the approximate spectral range from
which the RV signal is drawn.

All together there were 18 adjustable parameters, of which 12 were
controlled almost exclusively by the RV data, and 6 by the photometric
data. The data set had 105 RVs and 107 flux data points. Thus, the
total number of degrees of freedom was 194, of which 93 pertained to
RVs and 101 to photometry.

We determined the best-fitting parameter values and their 68.3\%
confidence limits with a Monte Carlo Markov Chain algorithm that we
have described elsewhere (see, e.g., Winn et al.~2007). Uniform
priors were adopted for all parameters except for b's time of transit
and orbital period, for which we adopted Gaussian priors based on the
ephemeris of Bakos et al.~(2009). We doubled the quoted errors in the
ephemeris, out of concern that systematic errors or transit-timing
variations have affected the results. The likelihood was taken to be
$\exp(-\chi^2/2)$ with
\begin{equation}
\chi^2 =
\sum_{i=1}^{105}
  \left[ \frac{v_i( \textnormal{obs} ) - v_i( \textnormal{calc} )}{\sigma_i} \right]^2 +
\sum_{j=1}^{107}
  \left[ \frac{f_j( \textnormal{obs} ) - f_j( \textnormal{calc} )}{\sigma_j} \right]^2 =
\chi^2_v + 
\chi^2_f,
\end{equation}
where $v_i({\rm obs})$ and $\sigma_i$ are the RV data and associated
uncertainties, $v_i({\rm calc})$ are the calculated RVs; $f_j({\rm
  obs})$ and $\sigma_j$ are the flux data and associated
uncertainties; and $f_j({\rm calc})$ are the calculated fluxes.

Each flux uncertainty $\sigma_i$ was taken to be the scatter in the
$\approx$16 data points contributing to each 4~min time bin. Each RV
uncertainty $\sigma_j$ was taken to be the quadrature sum of the
internally-estimated measurement error and a ``jitter'' term of
3.4~m~s$^{-1}$. The jitter term was set by the requirement $\chi^2_v =
93$, the relevant number of degrees of freedom, and is consistent with
the empirical jitter estimates of Wright~(2005) for stars similar to
HAT-P-13. In the best-fitting model, $\chi^2_f = 109.5$ and $\chi^2_v
= 93.0$, the rms photometric residual was 470~ppm and the rms RV
residual was 3.6~m~s$^{-1}$.

\begin{figure}[h]
\epsscale{1.0}
\plotone{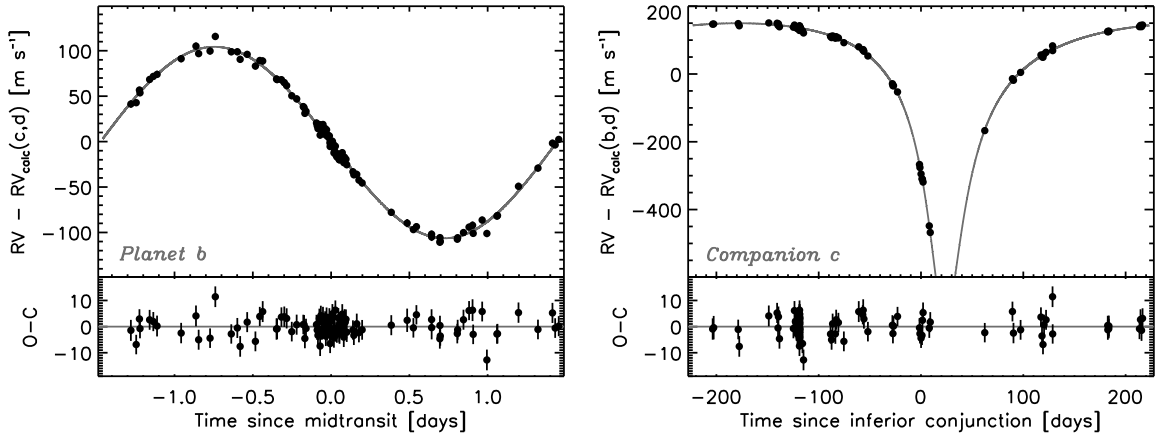}
\caption{ {\bf Radial-velocity variation as a function of orbital
    phase.} {\it Left.}---RV variation as a function of the orbital
  phase of planet b after subtracting the calculated variation due to
  c and d. {\it Right.}---RV variation as a function of the orbital
  phase of c, after subtracting the calculated variation due to b and
  d. \label{fig:rvphase}}
\end{figure}

\begin{figure}[htpb]
\epsscale{0.8}
\plotone{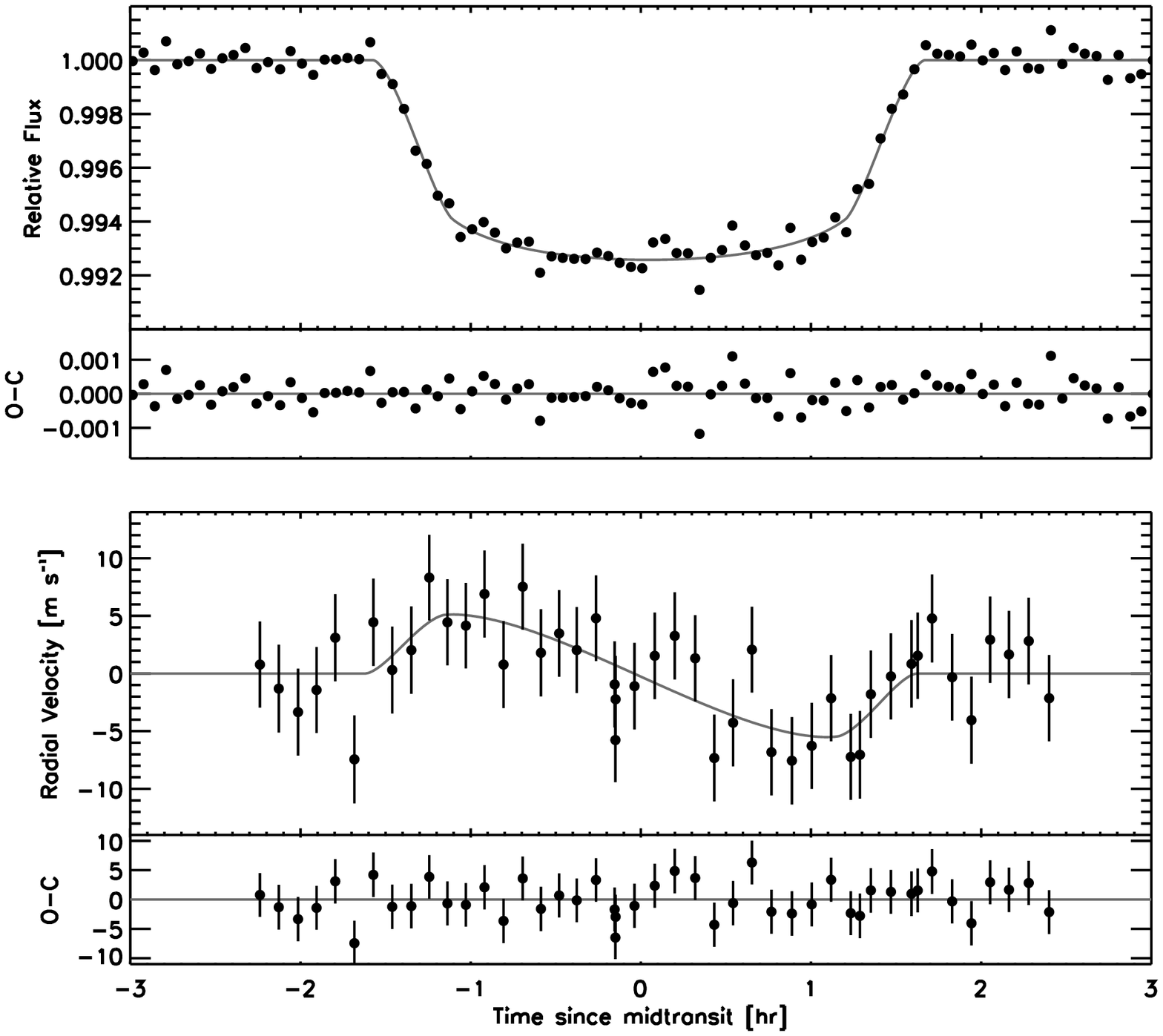}
\caption{ {\bf Transit photometry and radial-velocity variation.}
{\it Top.}---A composite transit light curve based on the $i'$-band
photometric data of Bakos et al.~(2009). Also plotted are the best-fitting
model and the residuals.
{\it Bottom}.---The apparent RV variation observed during
the transit phase, after subtracting the calculated contributions
due to orbital motion. The observed variation is interpreted as
the anomalous velocity due to the Rossiter-McLaughlin effect.
\label{fig:transit}}
\end{figure}

Table~\ref{tbl:params} gives the results for the parameter
values. Figure~\ref{fig:rvphase} shows the RVs as a function of the
orbital phases of b and c, expressed in days. In the left panel, the
RVs are plotted as a function of the orbital phase of b, after
subtracting the calculated contributions to the RV from companions c
and d. (The contribution due to d is a linear function of time.)
Likewise, the right panel of Figure~\ref{fig:rvphase} shows the RVs as
a function of the orbital phase of c, after subtracting the calculated
contributions from b and d. Figure~\ref{fig:transit} shows the data
over a restricted time range centered on b's transit. The top panel
shows the light curve. The bottom panel shows the data after
subtracting the calculated orbital RV, thereby isolating the RM
effect.

\subsection{Evidence for a third companion}
\label{subsec:third}

The extra acceleration, $\dot{\gamma}$, was included in the RV model
because a model consisting of only two Keplerian orbits gave an
unacceptable fit to the data.  With $\dot{\gamma} = 0$, the
RV-specific portions of the data and model had $\chi^2_v = 458.6$ and
94 degrees of freedom ($\chi^2_v/N_{{\rm dof}, v} = 4.9)$. The pattern
of residuals is displayed in Figure~\ref{fig:rvtime1}. There are large
and time-correlated residuals that are not easily attributed to
stellar jitter or underestimated measurement errors.

In contrast, when $\dot{\gamma}$ was allowed to vary freely, the
best-fitting model had $\dot{\gamma} = 17.5$~m~s$^{-1}$~yr$^{-1}$, and
$\chi^2_v=93$ with 93 degrees of freedom. The exact match between
$\chi^2_v$ and $N_{{\rm dof},v}$ is not significant in itself, as it
follows from our choice of 3.4~m~s$^{-1}$ for the jitter term.
However, it is significant that an acceptable fit was found for a
choice of jitter term that is in line with observations of similar
stars. Even more significant is that the correlated pattern of
residuals vanished. As shown in Figure~\ref{fig:rvtime2}, the
residuals scattered randomly around zero.

The failure of the two-Keplerian model, and the success of a model
with an additional radial acceleration, is evidence for a third
companion to HAT-P-13 (``d'') with a long orbital period. With the
limited information available, though, the properties of d are largely
unknown. Assuming its orbit to be nearly circular, and its mass to be
much smaller than that of the star, we may set $\dot{\gamma} \sim
GM_d\sin i_d/a_d^2$ to give an order-of-magnitude constraint
\begin{equation}
\left(\frac{M_d\sin i_d}{M_{\rm Jup}}\right)
\left(\frac{a_d}{{\rm 10~AU}} \right)^{-2} \sim 9.8,
\label{eq:accel}
\end{equation}
where $a_d$ is the orbital distance. By this standard, the newly
discovered object could be a 2.5~$M_{\rm Jup}$ planet at 5~AU, or a
10~$M_{\rm Jup}$ planet at 10~AU, or a 90~$M_{\rm Jup}$
(0.09~$M_\odot$) star at 30~AU, etc. The properties of d could be
substantially different depending on its eccentricity, argument of
pericenter, and time of conjunction. Orbits closer than $\sim$5~AU
would be subject to additional constraints by the requirement of
dynamical stability.

More information about d could be gleaned from any significant {\it
  curvature} in the RV signal, beyond the effects of the two Keplerian
orbits and a linear trend. We experimented with models that include a
``jerk'' parameter, $\ddot{\gamma}$, finding this parameter to be
highly covariant with the mass, orbital period, and eccentricity of
c. More elaborate models and detailed constraints on companion d will
only be justified after another few years of observing, when the
properties of companion c will have been well established. The
uncertainties given in Table~\ref{tbl:params} must therefore be
understood as subject to the assumption that d is producing no
significant RV curvature.

\subsection{Orbital eccentricities}
\label{subsec:ecc}

\begin{figure}[h]
\epsscale{1.0}
\plotone{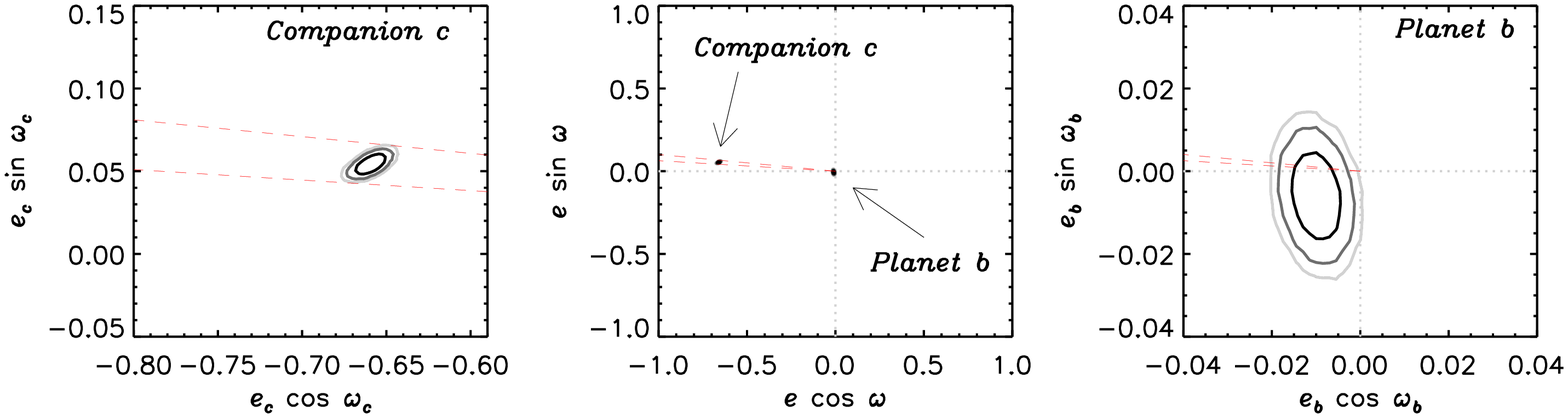}
\caption{ {\bf Results for the orbital eccentricities.}  The middle
  panel displays the results for b and c, while the left and right
  panels zoom in on the results each object. The contours enclose
  68\%, 95\%, and 99.73\% of the MCMC samples. The dashed lines show
  the 99.73\% confidence range for the apsidal orientation of c's
  orbit; they allow a visual assessment of the degree of apsidal
  alignment, and show that the limiting uncertainty in $\Delta\omega$
  is the large uncertainty in $e_b\sin\omega_b$. \label{fig:ecc}}
\end{figure}

Figure~\ref{fig:ecc} shows the results for the orbital eccentricities.
Planet c's orbit is strongly eccentric, with $e_c=0.6616\pm 0.0054$.
Planet b's orbit is nearly circular, with $e_b=0.0133\pm 0.0041$. To
assess the significance of the detection of eccentricity, it is
simpler to examine the components of the eccentricity vector because
they obey Gaussian distributions, while $e$ obeys a Rayleigh
distribution (see, e.g., Shen \& Turner 2008). We found
$e_b\cos\omega_b = -0.0099\pm 0.0036$ (i.e., nonzero with 2.8$\sigma$
significance) and $e_c\sin\omega_c=-0.0060\pm 0.0069$. The
eccentricity of b's orbit is right on the edge of
detectability.\footnote{For this reason the results are also sensitive
  to the choice of priors for the fitting parameters. The results
  described in this section and given in Table~\ref{tbl:params} are
  based on uniform priors for $e_b\cos\omega_b$ and
  $e_b\sin\omega_b$. If instead uniform priors are adopted for $e_b$
  and $\omega_b$, then we find $e_b = 0.0119 \pm 0.0040$.}

Because of the low significance of this detection, it is impossible to
draw a firm conclusion about whether its orbit is aligned with that of
companion c. Our result is $\Delta\omega \equiv \omega_b-\omega_c =
36_{-36}^{+27}$~degrees. The red dashed lines in Figure~\ref{fig:ecc}
show the 3$\sigma$ allowed region for $\omega_c$. The lines intersect
the allowed region for planet b, as shown in the right panel. Most of
the uncertainty in $\Delta\omega$ arises from the poorly determined
orientation of b's orbit.\footnote{If we {\it assume} the orbits are
  apsidally locked, and repeat the fitting procedure with the
  requirement $\Delta\omega=0$, we find $e_b=0.0104\pm 0.0032$.}

The best way to check on the eccentricity of b's orbit would be to
observe an occultation with the {\it Spitzer Space Telescope}. The
timing of occultations would allow $e_b\cos\omega_b$ to be determined
with an accuracy of about 0.001, several times better than the RV
result.  However, even after such an observation, considerable
uncertainty would remain in $\Delta\omega$ because the accuracy in
$e_b\sin\omega_b$ would not be much improved.

\subsection{The Rossiter-McLaughlin effect}
\label{subsec:lambda}

\begin{figure}[ht]
\epsscale{0.65}
\plotone{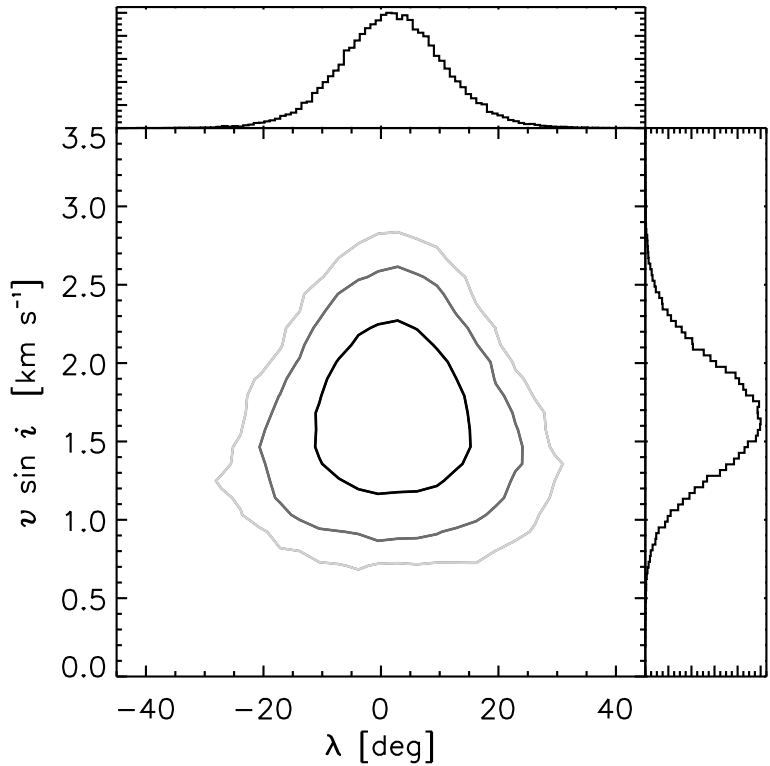}
\caption{ {\bf Results for the Rossiter-McLaughlin parameters,}
  based on our MCMC
  analysis of the RV data. The contours represent 68\%, 95\%, and
  99.73\% confidence limits, and the one-dimensional (marginalized)
  posterior probability distributions are shown on the sides of the
  contour plot.
  \label{fig:lambda}}
\end{figure}

The RV data obtained during transits exhibit a prograde
Rossiter-McLaughlin effect: an anomalous redshift for the first half
of the transit, followed by an anomalous blueshift for the second
half. The fit to the data is shown in Figure~\ref{fig:transit}, and
the resulting constraints on $\lambda$ and $v\sin i_\star$ are shown
in Figure~\ref{fig:lambda}. The finding of $\lambda=1.9\pm 8.6$~deg
implies a close alignment between the rotational angular momentum of
the star, and the orbital angular momentum of the planet, at least as
projected on the sky. Our result for the projected stellar rotation
velocity, $v\sin i_\star=1.66\pm 0.37$~km~s$^{-1}$, is about 1$\sigma$
smaller than the result of $2.9\pm 1.0$~km~s$^{-1}$ reported by Bakos
et al.~(2009).

\subsection{Inferior conjunction of planet c}
\label{subsec:dt}

It is not yet known whether the inclination of c's orbit is close
enough to 90$^\circ$ for transits to occur. Observations of transits
would reveal the mass and radius of the companion, allow a more
precise characterization of its orbit, and place constraints on the
mutual inclination of orbits b and c.

\begin{figure}[ht]
\epsscale{0.65}
\plotone{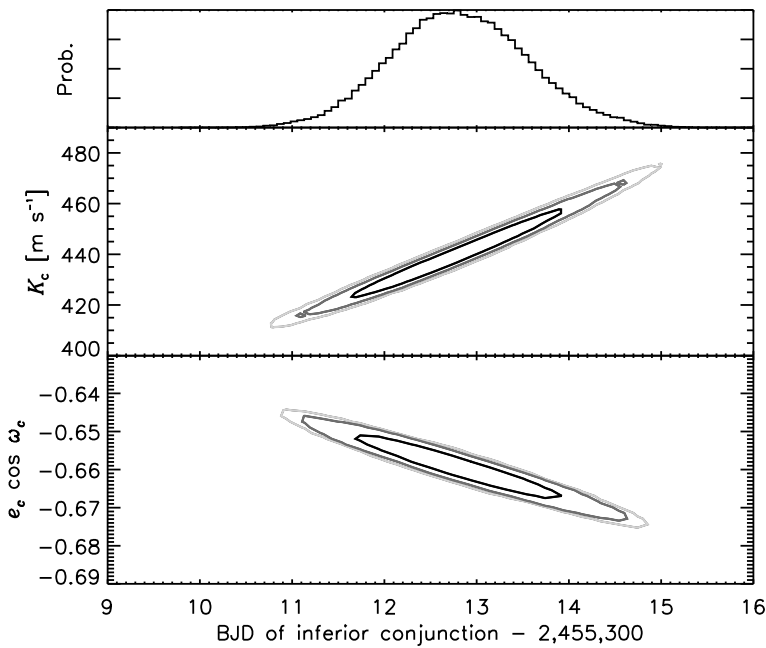}
\caption{ {\bf Results for the timing of the inferior
  conjunction of companion c,} based on our MCMC
  analysis of the RV data. The top panel shows
  the one-dimensional (marginalized)
  posterior probability distribution. The lower two panels
  illustrate the correlation with the other poorly-determined
  parameters of companion c. The contours represent 68\%, 95\%, and
  99.73\% confidence limits.
  \label{fig:dt}}
\end{figure}

Using our results we predicted the times of inferior conjunctions of
planet c, which is when transits would occur. The accuracy of the
predicted time is limited by correlations with the uncertainties in
c's velocity semiamplitude and eccentricity (see
Figure~\ref{fig:dt}). Table~\ref{tbl:prediction} gives the
results. The quoted uncertainties represent 1$\sigma$ (68.3\%)
confidence levels. It would be prudent to keep the star under
continuous photometric surveillance for the entire $3\sigma$ time
range, at least. The maximum transit duration is 14.9~hr.

\section{Discussion}
\label{sec:disc}

HAT-P-13 was already a noteworthy system, as the first known case of a
star with a transiting planet and a second close companion.  We have
presented evidence for a third companion in the form of a long-term
radial acceleration of the star. The properties of the newly
discovered long-period companion will remain poorly known until
additional RV data are gathered over a significant fraction of its
orbital period. Our analysis of the Rossiter-McLaughlin effect shows
that planet b's orbital axis is aligned with the stellar rotation
axis, as projected on the sky. Our new data also agree with the
previous finding that the orbit of planet b is slightly eccentric.

The latter two findings are relevant to the second reason why HAT-P-13
is noteworthy: its orbital configuration may represent an example of
two-planet tidal evolution. In this scenario, first envisioned by Wu
\& Goldreich (2002) and investigated further by Mardling~(2007), tidal
circularization of the inner planet's orbit is delayed due to
gravitational interactions with the outer planet. The interactions
drive the system into a state of apsidal alignment, where it remains
as both orbits are slowly circularized. As it turned out, the specific
planetary system that inspired Wu \& Goldreich~(2002) was irrelevant
to their theory, because the ``outer planet'' was found to be a
spurious detection (Butler et al.~2002).

Batygin et al.~(2009) welcomed HAT-P-13 as a genuine system that
followed the path predicted by Wu \& Goldreich~(2002), and with the
additional virtue that the inner planet is transiting. For this
interpretation to be valid, the apsides of b and c must be aligned,
whereas we have found the angle between the apsides to be
$36_{-36}^{+27}$~deg, differing from zero by 1$\sigma$. We do not
consider this result to be significant enough to draw a firm
conclusion, especially in light of the uncertainties due to the {\it
  ad hoc} stellar jitter term and our simplified treatment of the
influence of companion d. Further RV monitoring and observations of
occultations are needed to make progress.

Batygin et al.~(2009) also showed that the existence of transits would
allow for an empirical estimate of the tidal Love number $k_2$ of
planet b, as mentioned in \S~1. The requirement that the apsidal
precession rates of b and c are equal leads to a condition on $k_2$,
because b's precession rate is significantly affected by its tidal
bulge. Subsequent work by Mardling~(2010) showed that for a unique
determination of $k_2$ it is necessary for the mutual inclination
$\Delta i$ between orbits b and c to be small. If instead the orbits
are mutually inclined, then tidal evolution drives the system into a
state in which $e_b$ and $\Delta i$ undergo oscillations: a cycle in
parameter space, instead of a fixed point. Furthermore,
Mardling~(2010) argued that a large mutual inclination should be
considered plausible, or even likely, given c's high eccentricity. She
proposed that b and c began with nearly circular and coplanar orbits,
but c's orbit was strongly perturbed by an interaction with a
hypothetical outer planet. Those same perturbations would likely have
tilted c's orbit.

The relation, if any, between the newly-discovered HAT-P-13d and
Mardling's hypothetical outer planet is unclear. In her scenario, the
outer planet is ejected from the system. This seems important to the
scenario, as otherwise d would continue interacting with c, and
interfere with the tidal evolution of b and c. Thus, unless d's
pericenter was somehow raised to avoid further encounters with c, it
does not seem likely to have played the role envisioned by
Mardling~(2010). Of course the scenario could still be correct even if
the third companion d was not the scattering agent; a fourth (ejected)
companion may have been responsible.

Our study of the Rossiter-McLaughlin effect pertains to the angle
$\psi_{\star,b}$ between planet b's orbit and the stellar equator, and
has no {\it direct} bearing on the angle $\Delta i$ between the
orbital planes of b and c. However, there is an {\it indirect}
connection, through the nodal precession that would be caused by
mutually inclined orbits.  As shown by Mardling~(2010), planet b is
far enough from the star that its orbital precession rate is likely to
be dominated by the torque from c, rather than the quadrupole moment
$J_2$ of the star. The critical orbital distance inside which the
stellar torque is dominant is $\sim$$(2 J_2 a_c^2 M_c/M_\star)^{0.2}$
(Burns 1986), which is 0.020~AU assuming $J_2=2\times 10^{-7}$ as for
the Sun. This is smaller than the actual orbital distance of 0.043~AU.
Hence if $\Delta i$ were large, then b's orbit would nodally precess
around c's orbital axis, which would cause periodic variations in
$\psi_{\star,b}$. Therefore, at any given moment in the system's
history, we would be unlikely to observe a small value of
$\psi_{\star,b}$ unless $\Delta i$ were small. However, it is
impossible to draw firm conclusions about $\Delta i$ because of the
dependence on initial conditions, the possible effects of companion d,
and the fact that only the sky-projected angle $\lambda$ is measured
rather than the true obliquity $\psi_{\star,b}$.

It may be possible to estimate $\Delta i$ based on transit timing
variations of planet b (Nesvorn{\'y} \& Beaug{\'e} 2010; Bakos et
al.~2009). An even more direct estimate of $\Delta i$ could be
achieved if transits of c were detected. The existence of transits
would show that $i_c$ is nearly $90^\circ$, as is $i_b$. This would
suggest $\Delta i$ is small, although it would still be possible that
the orbits are misaligned and their line of nodes happens to lie along
the line of sight. The most definitive result would be obtained by
observing the Rossiter-McLaughlin effect during transits of c, and
comparing c's value of $\lambda$ with that of planet b. In effect, the
rotation axis of the star would be used as a reference line from which
the orientation of each orbit is measured (Fabrycky 2009). This gives
additional motivation to observe HAT-P-13 throughout the upcoming
conjunctions of companion c.

\acknowledgments We thank Dan Fabrycky for helpful conversations,
especially about the dynamical implications of our results. We also
thank Debra Fischer and John Brewer for investigating the
spectroscopic determination of the stellar rotation rate. We are
grateful to Scott Gaudi and Greg Laughlin for comments on the
manuscript.

J.N.W.\ gratefully acknowledges support from the NASA Origins program
through award NNX09AD36G and the MIT Class of 1942. A.W.H.\
acknowledges a Townes Postdoctoral Fellowship from the Space Sciences
Laboratory at UC Berkeley. G.A.B.\ was supported by NASA grant
NNX08AF23G and an NSF Astronomy \& Astrophysics Postdoctoral
Fellowship (AST-0702843). G.T.\ acknowledges partial support from NASA
grant NNX09AF59G. S.A.\ acknowledges the support of the Netherlands
Organisation for Scientific Research (NWO). N.N.\ was supported by a
Japan Society for Promotion of Science (JSPS) Fellowship for Research
(PD:~20-8141). J.N.W.\ and N.N.\ were also supported in part by the
National Science Foundation under Grant No.\ NSF PHY05-51164 (KITP
program ``The Theory and Observation of Exoplanets'' at UCSB).

The data presented herein were obtained at the W.M.~Keck Observatory,
which is operated as a scientific partnership among the California
Institute of Technology, the University of California, and the
National Aeronautics and Space Administration, and was made possible
by the generous financial support of the W.M.~Keck Foundation. We
extend special thanks to those of Hawaiian ancestry on whose sacred
mountain of Mauna Kea we are privileged to be guests. Without their
generous hospitality, the Keck observations presented herein would not
have been possible.

{\it Facilities:} \facility{Keck:I (HIRES)}

\begin{deluxetable}{lccr}

\tabletypesize{\tiny}
\tablecaption{Relative Radial Velocity Measurements of HAT-P-13\label{tbl:rv}}
\tablewidth{0pt}

\tablehead{
\colhead{BJD} &
\colhead{RV [m~s$^{-1}$]} &
\colhead{Error [m~s$^{-1}$]}
}

\startdata
  $  2454548.80650$  &  $     87.29$  &  $   2.00$  \\
  $  2454548.90850$  &  $     70.55$  &  $   1.44$  \\
  $  2454602.73396$  &  $    -77.76$  &  $   1.49$  \\
  $  2454602.84691$  &  $    -77.84$  &  $   1.72$  \\
  $  2454603.73415$  &  $     82.29$  &  $   1.41$  \\
  $  2454603.84324$  &  $    102.65$  &  $   2.05$  \\
  $  2454633.77241$  &  $    112.70$  &  $   2.00$  \\
  $  2454634.75907$  &  $    -57.09$  &  $   1.97$  \\
  $  2454635.75475$  &  $     86.55$  &  $   2.12$  \\
  $  2454636.74969$  &  $    107.21$  &  $   1.80$  \\
  $  2454727.13850$  &  $    117.62$  &  $   1.90$  \\
  $  2454728.13189$  &  $    -58.37$  &  $   1.66$  \\
  $  2454778.07301$  &  $    -57.70$  &  $   1.40$  \\
  $  2454779.08373$  &  $    120.17$  &  $   1.71$  \\
  $  2454780.09368$  &  $    -13.75$  &  $   1.88$  \\
  $  2454791.11129$  &  $     92.67$  &  $   1.64$  \\
  $  2454809.99575$  &  $   -114.15$  &  $   2.39$  \\
  $  2454839.06085$  &  $   -225.51$  &  $   1.54$  \\
  $  2454865.02660$  &  $   -448.40$  &  $   1.49$  \\
  $  2454867.90311$  &  $   -488.00$  &  $   2.88$  \\
  $  2454928.83635$  &  $   -289.10$  &  $   1.44$  \\
  $  2454955.86964$  &  $   -186.54$  &  $   1.63$  \\
  $  2454956.86327$  &  $     -5.48$  &  $   1.90$  \\
  $  2454963.85163$  &  $   -119.86$  &  $   1.62$  \\
  $  2454983.74976$  &  $     41.30$  &  $   1.50$  \\
  $  2454984.76460$  &  $   -134.63$  &  $   1.51$  \\
  $  2454985.73856$  &  $     19.77$  &  $   1.50$  \\
  $  2454986.76358$  &  $     25.83$  &  $   1.75$  \\
  $  2454988.74066$  &  $     50.77$  &  $   1.68$  \\
  $  2455109.11745$  &  $    143.40$  &  $   2.24$  \\
  $  2455110.10818$  &  $    -30.20$  &  $   2.91$  \\
  $  2455134.11719$  &  $     48.56$  &  $   1.55$  \\
  $  2455135.13125$  &  $    168.31$  &  $   1.97$  \\
  $  2455164.01155$  &  $    181.07$  &  $   2.06$  \\
  $  2455172.12118$  &  $     72.55$  &  $   1.78$  \\
  $  2455173.02454$  &  $    172.95$  &  $   1.73$  \\
  $  2455188.04447$  &  $    102.62$  &  $   1.53$  \\
  $  2455189.08587$  &  $    -25.28$  &  $   1.29$  \\
  $  2455189.98539$  &  $    140.26$  &  $   1.30$  \\
  $  2455191.11450$  &  $     67.47$  &  $   1.49$  \\
  $  2455192.02847$  &  $    -33.41$  &  $   1.50$  \\
  $  2455193.85943$  &  $    105.31$  &  $   1.49$  \\
  $  2455193.86475$  &  $    104.10$  &  $   1.48$  \\
  $  2455193.86961$  &  $     97.82$  &  $   1.49$  \\
  $  2455193.94390$  &  $     87.20$  &  $   1.50$  \\
  $  2455193.94850$  &  $     84.10$  &  $   1.68$  \\
  $  2455193.95323$  &  $     81.03$  &  $   1.59$  \\
  $  2455193.95775$  &  $     81.95$  &  $   1.49$  \\
  $  2455193.96234$  &  $     85.47$  &  $   1.61$  \\
  $  2455193.96702$  &  $     73.88$  &  $   1.69$  \\
  $  2455193.97167$  &  $     84.75$  &  $   1.63$  \\
  $  2455193.97628$  &  $     79.58$  &  $   1.59$  \\
  $  2455193.98097$  &  $     80.27$  &  $   1.64$  \\
  $  2455193.98536$  &  $     85.58$  &  $   1.44$  \\
  $  2455193.98980$  &  $     80.71$  &  $   1.51$  \\
  $  2455193.99433$  &  $     79.41$  &  $   1.42$  \\
  $  2455193.99888$  &  $     81.14$  &  $   1.59$  \\
  $  2455194.00354$  &  $     73.98$  &  $   1.60$  \\
  $  2455194.00825$  &  $     79.68$  &  $   1.49$  \\
  $  2455194.01271$  &  $     72.95$  &  $   1.60$  \\
  $  2455194.01716$  &  $     73.63$  &  $   1.54$  \\
  $  2455194.02147$  &  $     71.22$  &  $   1.49$  \\
  $  2455194.02618$  &  $     72.92$  &  $   1.45$  \\
  $  2455194.03098$  &  $     64.82$  &  $   1.55$  \\
  $  2455194.03561$  &  $     64.92$  &  $   1.54$  \\
  $  2455194.04057$  &  $     66.44$  &  $   1.55$  \\
  $  2455194.04546$  &  $     67.07$  &  $   1.62$  \\
  $  2455194.05048$  &  $     64.00$  &  $   1.51$  \\
  $  2455194.05515$  &  $     54.30$  &  $   1.56$  \\
  $  2455194.05976$  &  $     56.32$  &  $   1.60$  \\
  $  2455194.06437$  &  $     61.63$  &  $   1.46$  \\
  $  2455194.06915$  &  $     51.66$  &  $   1.52$  \\
  $  2455194.07416$  &  $     49.80$  &  $   1.63$  \\
  $  2455194.07901$  &  $     50.00$  &  $   1.50$  \\
  $  2455194.08376$  &  $     53.07$  &  $   1.52$  \\
  $  2455194.08858$  &  $     46.90$  &  $   1.47$  \\
  $  2455194.09350$  &  $     51.24$  &  $   1.63$  \\
  $  2455194.09842$  &  $     51.69$  &  $   1.50$  \\
  $  2455194.10345$  &  $     51.65$  &  $   1.65$  \\
  $  2455194.10848$  &  $     54.47$  &  $   1.68$  \\
  $  2455194.11341$  &  $     48.28$  &  $   1.54$  \\
  $  2455194.11813$  &  $     43.50$  &  $   1.59$  \\
  $  2455194.12270$  &  $     49.47$  &  $   1.51$  \\
  $  2455194.12732$  &  $     47.17$  &  $   1.63$  \\
  $  2455194.13216$  &  $     47.26$  &  $   1.57$  \\
  $  2455194.13716$  &  $     41.20$  &  $   1.51$  \\
  $  2455194.17667$  &  $     33.41$  &  $   1.44$  \\
  $  2455196.94719$  &  $     59.88$  &  $   1.29$  \\
  $  2455197.94842$  &  $    -36.39$  &  $   2.02$  \\
  $  2455229.08581$  &  $     20.75$  &  $   1.66$  \\
  $  2455229.87780$  &  $    -60.13$  &  $   1.73$  \\
  $  2455232.01621$  &  $     22.91$  &  $   1.52$  \\
  $  2455251.92524$  &  $     96.71$  &  $   2.09$  \\
  $  2455255.82341$  &  $    -93.36$  &  $   1.40$  \\
  $  2455256.97046$  &  $     54.58$  &  $   1.47$  \\
  $  2455260.85979$  &  $     39.52$  &  $   1.54$  \\
  $  2455284.82534$  &  $   -171.99$  &  $   1.73$  \\
  $  2455285.89491$  &  $    -97.21$  &  $   1.75$  \\
  $  2455289.81794$  &  $    -32.19$  &  $   1.31$  \\
  $  2455311.74995$  &  $   -418.30$  &  $   1.56$  \\
  $  2455312.83027$  &  $   -254.42$  &  $   1.53$  \\
  $  2455313.74879$  &  $   -405.93$  &  $   1.39$  \\
  $  2455314.80031$  &  $   -436.91$  &  $   1.87$  \\
  $  2455320.86712$  &  $   -514.39$  &  $   1.58$  \\
  $  2455321.81620$  &  $   -436.55$  &  $   1.73$  
\enddata

\tablecomments{The RV was measured relative to an arbitrary template
  spectrum; only the differences are significant. The uncertainty
  given in Column 3 is the internal error only and does not account
  for ``stellar jitter.''}

\end{deluxetable}

\clearpage

\begin{deluxetable}{lc}
 
\tabletypesize{\footnotesize}
\tablecaption{Model Parameters for HAT-P-13\label{tbl:params}}
\tablewidth{0pt}
 
\tablehead{
\colhead{Parameter} &
\colhead{Value}
}

\startdata \hline
{\it Star} \\
 \hline
Mass, $M_\star$~[$M_\odot$]                              & $1.22^{+0.05}_{-0.10}$ \\
Radius, $R_\star$~[$R_\odot$]                            & $1.559 \pm 0.080$ \\
Projected stellar rotation rate, $v\sin i_\star$~[km~s$^{-1}$] & $1.66\pm 0.37$ \\
\hline
{\it Planet b} \\
\hline
Mass, $M_b$ [$M_{\rm Jup}$]                              & $0.851\pm 0.038$ \\
Radius, $R_b$ [$R_{\rm Jup}$]                            & $1.272\pm 0.065$ \\
Orbital period, $P_b$~[d]                               & $2.916250 \pm 0.000015$  \\
Planet-to-star radius ratio, $R_p/R_\star$              & $0.08389 \pm 0.00081$ \\
Star-to-orbit radius ratio, $R_\star/a$                 & $0.1697\pm 0.0072$ \\
Orbital inclination, $i$ [deg]                         & $83.40\pm 0.68$ \\
Impact parameter, $b_{\rm tra}$                         & $0.679\pm 0.043$ \\
Time of midtransit, $T_{{\rm tra},b}$~[BJD]              & $2,454,779.92976 \pm 0.00075$  \\
Orbital eccentricity, $e_b$                            & $0.0133\pm 0.0041$ \\
Argument of pericenter, $\omega_b$~[deg]               & $210_{-36}^{+27}$ \\
$e_b\cos\omega_b$                                      & $-0.0099\pm 0.0036$ \\
$e_b\sin\omega_b$                                      & $-0.0060\pm 0.0069$ \\
Velocity semiamplitude, $K_b$~[m~s$^{-1}$]            & $106.04 \pm 0.73$ \\
Projected spin-orbit angle, $\lambda$~[deg]            & $1.9\pm 8.6$ \\
\hline
{\it Companion c} \\
\hline
Minimum mass, $M_c\sin i_c$ [$M_{\rm Jup}$]             & $14.28\pm 0.28$ \\
Orbital period, $P_c$~[d]                              & $446.27\pm 0.22$  \\
Time of inferior conjunction, $T_{{\rm con},c}$~[BJD]    & $2,455,312.80 \pm 0.74$ \\
Orbital eccentricity, $e_c$                            & $0.6616\pm  0.0054$ \\
Argument of pericenter, $\omega_c$~[deg]               & $175.29\pm 0.35$ \\
$e_c\cos\omega_c$                                      & $-0.6594\pm 0.0056$ \\
$e_c\sin\omega_c$                                      & $0.0543\pm 0.0038$ \\
Velocity semiamplitude, $K_c$~[m~s$^{-1}$]              & $440 \pm 11$ \\
\hline
{\it Other system parameters} \\
\hline
Angle between apsides, $\omega_b-\omega_c$~[deg]    & $36_{-36}^{+27}$ \\
Velocity offset, $\gamma$ [m~s$^{-1}$]              & $-100.3 \pm 2.0$ \\
$\dot{\gamma}$ [m~s$^{-1}$~yr$^{-1}$]               & $17.51\pm 0.90$
\enddata

\end{deluxetable}

\begin{deluxetable}{lccccc}
 
\tabletypesize{\footnotesize}
\tablecaption{Predicted times of inferior conjunction for HAT-P-13c\label{tbl:prediction}}
\tablewidth{0pt}
 
\tablehead{
\colhead{Year} &
\colhead{Month} &
\colhead{Date} &
\colhead{Hour [UT]} &
\colhead{Julian Date} &
\colhead{Uncertainty [d]}
}

\startdata
 2010 & April &  26 &   7.3 &  2455312.80 &        0.74 \\
 2011 & July &  16 &  13.9 &   2455759.08 &        0.85 \\
 2012 & October &   4 &  20.4 &  2456205.35 &        1.00 \\
 2013 & December &  25 &   3.0 & 2456651.62 &        1.17 \\
 2015 & March &  16 &   9.6 &  2457097.90 &        1.36
\enddata

\end{deluxetable}

\end{document}